\documentclass[twocolumn,prl,aps,showpacs]{revtex4}
\usepackage{amsmath}
\usepackage{graphicx}
\usepackage{sistyle}

\begin{document}

\title{Single-photon space-like antibunching}

\author{Thiago Guerreiro$^1$, Bruno Sanguinetti$^1$, Hugo Zbinden$^1$ Nicolas Gisin$^1$, and Antoine Suarez $^2$}
\address{
$^1$Group of Applied Physics, University of Geneva,1211 Geneva 4, Switzerland
\\
$^2$Center for Quantum Philosophy, P.O.Box 304, 8044 Zurich,
Switzerland; suarez@leman.ch}

\date{April 5, 2012}

\begin{abstract}
We use heralded single photons to perform an antibunching experiment in which the clicks at the detectors are spacelike separated events. The idea of such experiment dates back to the 5th Solvay conference, when it was proposed by Einstein as an expression of his concerns about quantum theory.
\end{abstract}


\maketitle

Quantum theory is well known for being counterintuitive. Today, this is most often discussed in the context of the quantum nonlocality illustrated by the violation of Bell's inequality. However, historically, other counterintuitive aspects of quantum theory dominated, like the wave-particle duality, the Heisenberg uncertainty relations and the measurement problem. In this short article, we concentrate on a conundrum that triggered much of Einstein's suspicion towards quantum theory. 

During the famous 5th Solvay conference in 1927, Einstein considered a single particle which, after diffraction in a pin-hole encounters a ``detection plate'' (e.g. in the case of photons, a photographic plate), see Fig~1. We simplify this thought experiment, though keeping the essence, by replacing the ``detection plate'' by two detectors. Einstein noted that there is no question that only one of them can detect the particle, otherwise energy would not be conserved. However, he was deeply concerned about the situation in which the two detectors are space-like separated, as this prevents - according to relativity - any possible coordination among the detectors \cite{jammer}:
\\

\textit{``It seems to me,'' Einstein continued, ``that this difficulty cannot be overcome unless the description of the process in terms of the Schr\"{o}dinger wave is supplemented by some detailed specification of the localization of the particle during its propagation. I think M. de Broglie is right in searching in this direction.''}
\\

\begin{figure}[htbp]
\includegraphics[width=80 mm]{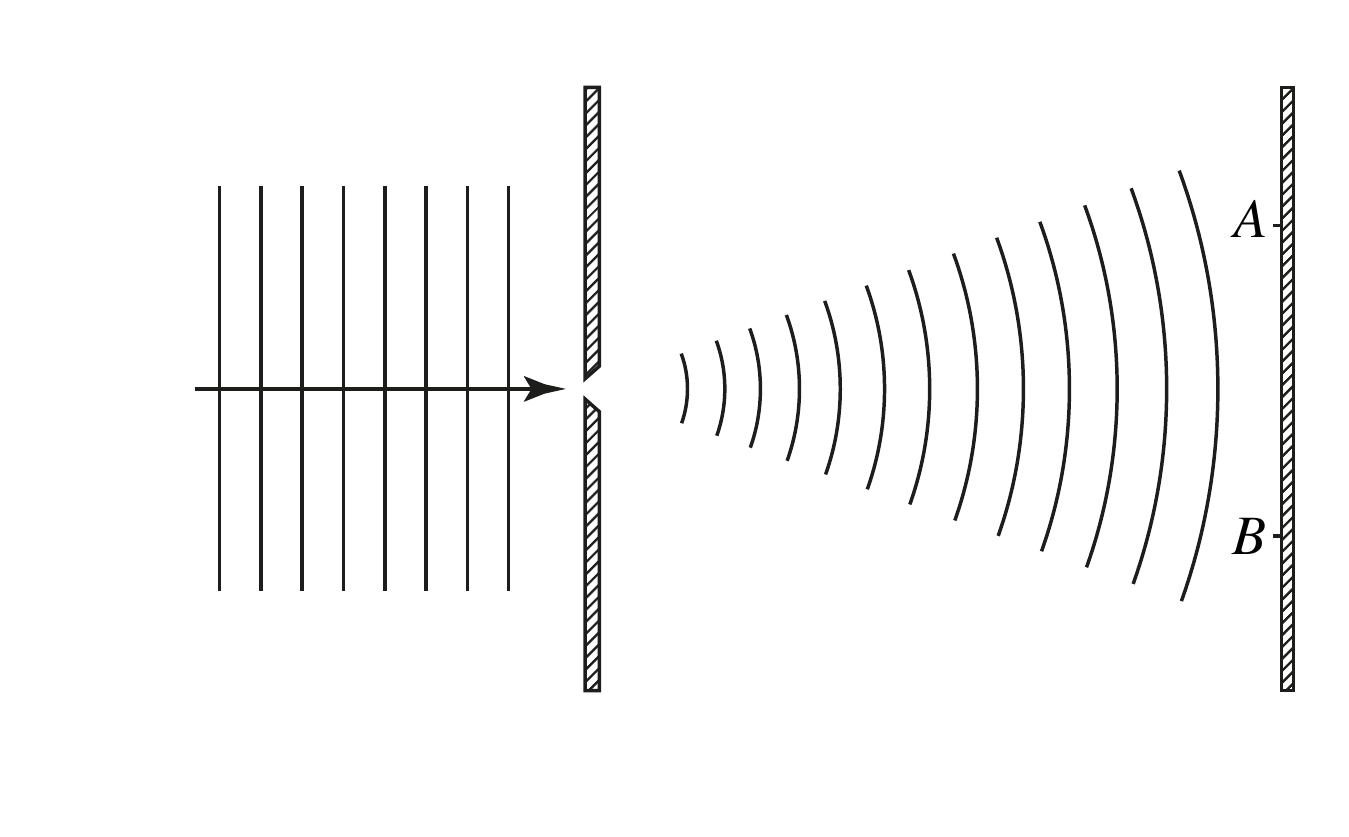}
\caption{Gedanken experiment proposed by Einstein during the 5th Solvey conference. A and B are points on the photographic plate, for which the events of detection can be spacelike separated from each other. Adapted from \cite{wheeler}.}
\label{f1}
\end{figure}
Yet, this simple argument got somehow buried in the Einstein-Bohr debates. Anyway, at the time there was plenty of new physics to explore and keep physicists busy; moreover, the technology didn't allow one to realize what was then only ``Gedanken experiments''. Nevertheless, Einstein's idea of ``supplementary specifications'' to de~Broglie's guiding wave ideas came back in 1952 in the form of an explicit model, presented by David Bohm \cite{bohm}. This model is empirically undistinguishable from quantum theory, as it makes precisely the same predictions. Applied to the above described situation, Bohm's model provides an elegant and simple solution to Einstein's conundrum: the particle follows one and only one continuous trajectory from the pin hole to one of the two detectors. Hence, which detector clicks is already determined at the pin-hole: the detector hit by the particle's trajectory detects it, while the other detector only ``sees'' an empty wave. 

Bohm's model applied to a single particle is manifestly local. However, when applied to two or more particles Bohm's model includes a nonlocal ``quantum potential''. This inspired John Bell to discover his famous inequality which, in turn, opened the fields of quantum nonlocality and more recently of device independent quantum information processing \cite{acin, bancal, pironio}, both very fruitful experimental and theoretical research fields.

\begin{figure*}[htbp]
\includegraphics[width=16cm]{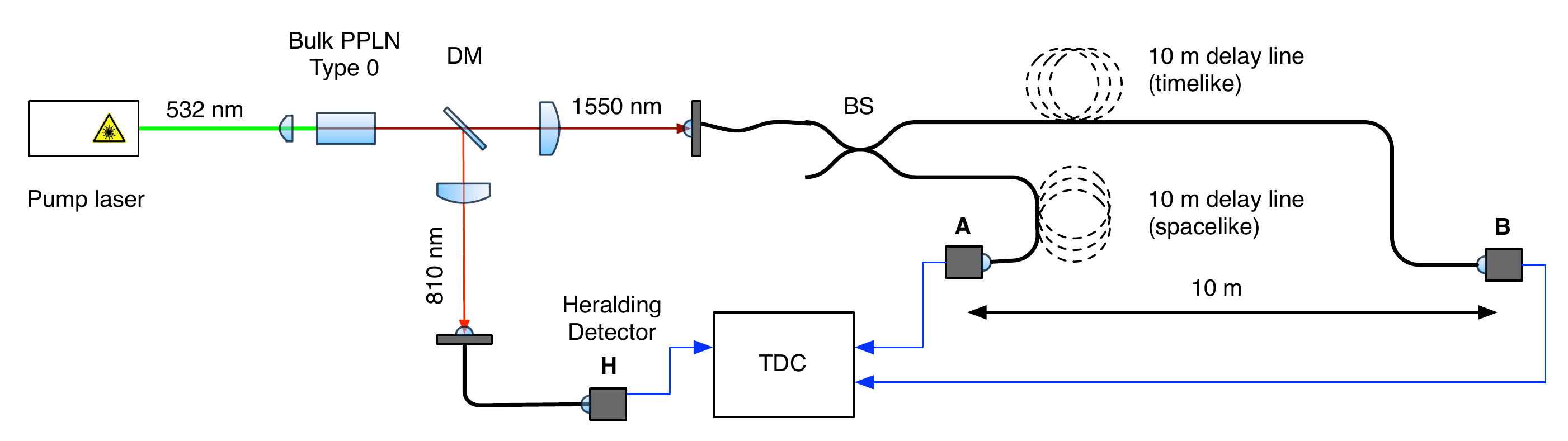}
\caption{Experimental setup: photon pairs are generated by Spontaneous Parametric Down Conversion at the wavelengths of \SI{1550}{nm} and \SI{810}{nm}. These pairs are split by a dichroic mirror (DM), and the 810 nm photon is sent to detector D, used to herald the presence of the \SI{1550}{nm} photon which follows to the beam splitter (BS). Arbitrary electronic delays were applied before TDC to ensure the coincidence peaks would remain on scale.}
\label{f2}
\end{figure*}

But what happened to Einstein's original ``Gedanken experiment''? Surprisingly, to the best of our knowledge, this simple - with today's technology - experiment had never been done, as Antoine Suarez taught us \cite{as10}. Producing correlated photon pair was first done using atomic cascades and used to test Bell's inequality \cite{clauser,aspect}. A bit later, this was used by Philippe Grangier to produce heralded single photons, one photon heralding the second one, and to perform beautiful experiments on single-photon interferences \cite{grangier}. Today, correlated photon pairs are routinely produced by spontaneous parametric downconversion in nonlinear crystals \cite{mandel}. This led to many demonstrations of heralded single photon sources, see e.g.~\cite{fasel}. 

Here we use one such source of heralded single photons and two single photon detectors to realize Einstein's Gedanken experiment. We take great care at assuring the space-like separation between the two detectors. We confirm that, in each round of the experiment, only one detector click. This is unsurprising but stresses the non local characteristic of the textbook collapse of the wave function at detection, even in single particle experiments.

\section{Experiment}

The experiment consists in verifying that when a single photon is thrown at a beam splitter, it is detected in only one arm, i.e. the probability $P_{AB}$ of getting a coincidence between the two detectors  $A$ and $B$ is much smaller than the product of the probabilities of detection on each side $P_{A}\times P_{B}$, as would be expected in the case of uncorrelated events.

The experimental setup is shown in Fig.~\ref{f2} and consists of a source of heralded single photons which is coupled into a single mode fiber and injected into a fiber beamsplitter (BS). Each of the two outputs of the beamsplitter goes to a single photon detector (IDQ ID200), detector \textbf{A} being close to the source and detector \textbf{B} being separated by a distance of approximately 10 meters.

If we ensure that the fiber lengths before each detector are equal by inserting a 10\,m (\SI{50}{ns}) fiber delay loop before detector \textbf{A}, the detections will happen simultaneously in some reference frame, thus being space-like separated (a signal would take 33\,ns to travel between the two detectors at the speed of light; simultaneity of detection is guaranteed to within 1\,ns by the matched length of fiber both before and inside the detectors). It is also possible to make the detections time-like separated by removing the 10\,m delay line from detector \textbf{A} and adding it to detector \textbf{B}.

\begin{figure}[htbp]
\includegraphics[width=80 mm]{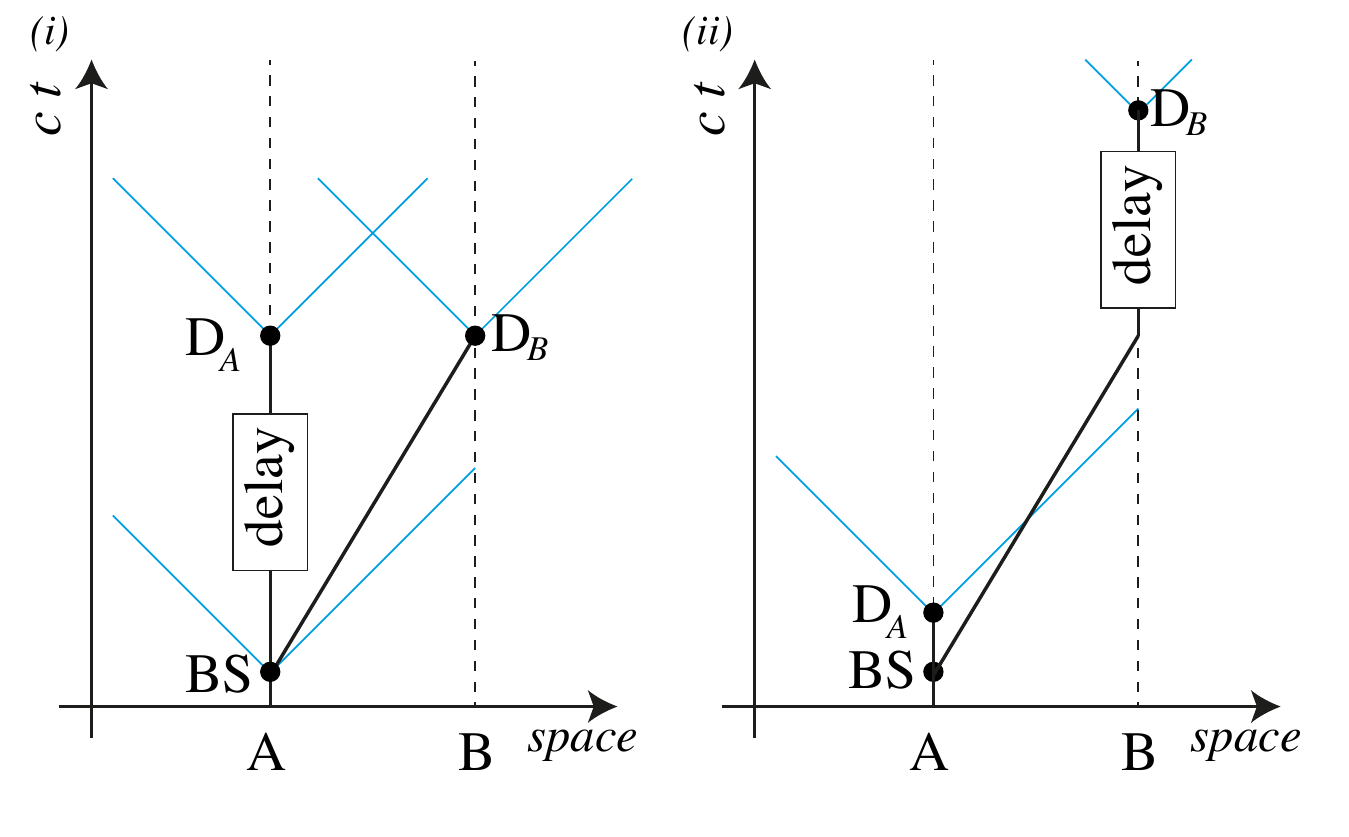}
\caption{Spacetime diagrams for spacelike (i) and timelike~(ii) configurations. A and B represent the locations of the detectors, with detector A being close to the beamsplitter~BS. $D_A$ and $D_B$ represent the detection events.}
\label{f1}
\end{figure}

The source consists of a periodically poled lithium niobate crystal pumped by a continuous wave laser at 532\,nm and emitting photon pairs with one photon at 810\,nm which is used to herald the other photon at 1550\,nm. The photons are coupled into single mode fiber with an efficiency of the order of 35\% and the detector efficiencies are of approximately 10\% for the 1550\,nm detectors (IDQ ID200) and 56\% for the 810\,nm detector (LaserComponents COUNT). 

Coincidences are taken with a Time to Digital Converter (TDC, Agilent CD890) which has a resolution of 50\,ps; we however use a coincidence window of 1\,ns as the detectors have a jitter of 400\,ps for the IDQs and \SI{1.2}{ns} for the Laser Components. Within this time the source has probability $p$ of generating one photon and probability $p_2\approx p^2/2$ of generating two photons. We lower the number of two photon events to a level where they have a rate comparable to the detector noise by reducing the pump power.

\section{Results}
\begin{figure}[htbp]
\includegraphics[width=80 mm]{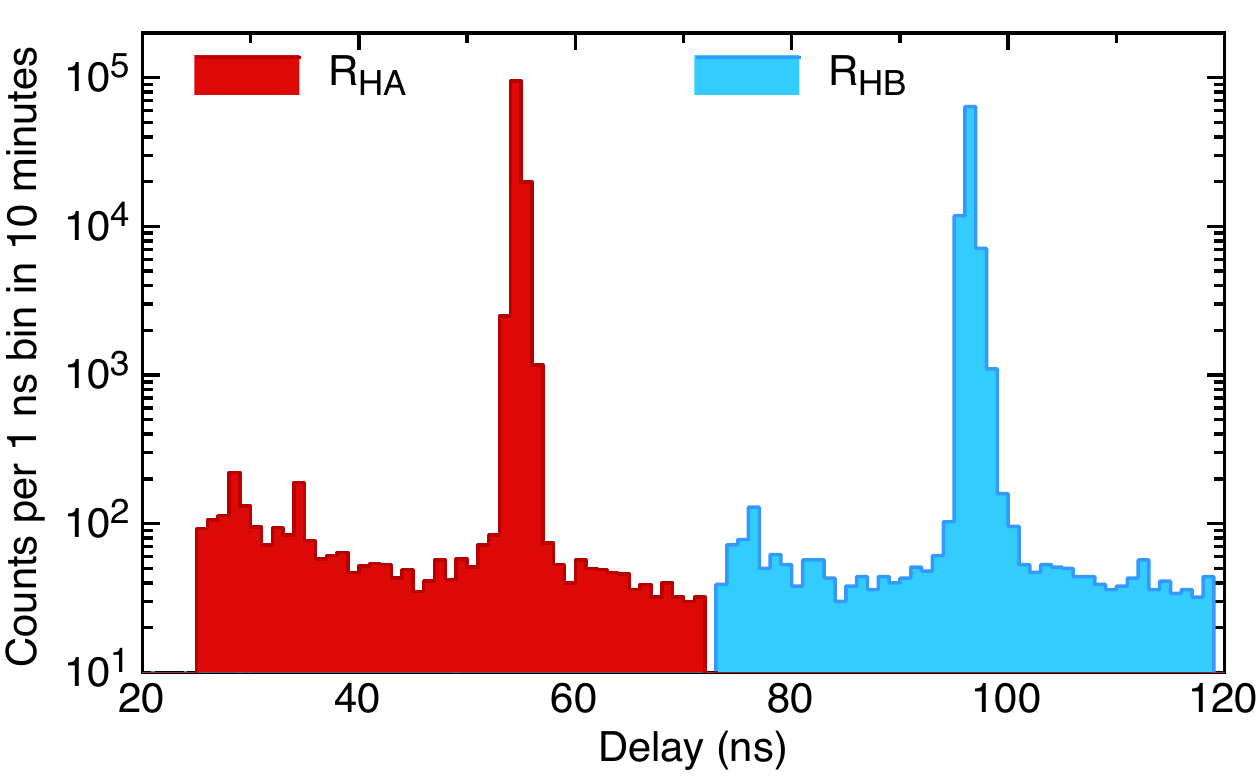}
\caption{Coincidences between the heralding detector and each of the detectors A (red) and B (blue) with spacelike separation, measured in a window of \SI{1}{ns} during a time period of 10 minutes. $R_{HA}$=\SI{9.49e4}/10\,min, $R_{HB}=$\SI{6.39e4}/10\,min. The noise is on average: $R_N=50$/10\,min.}
\label{f3}
\end{figure}

\begin{figure}[htbp]
\includegraphics[width=80 mm]{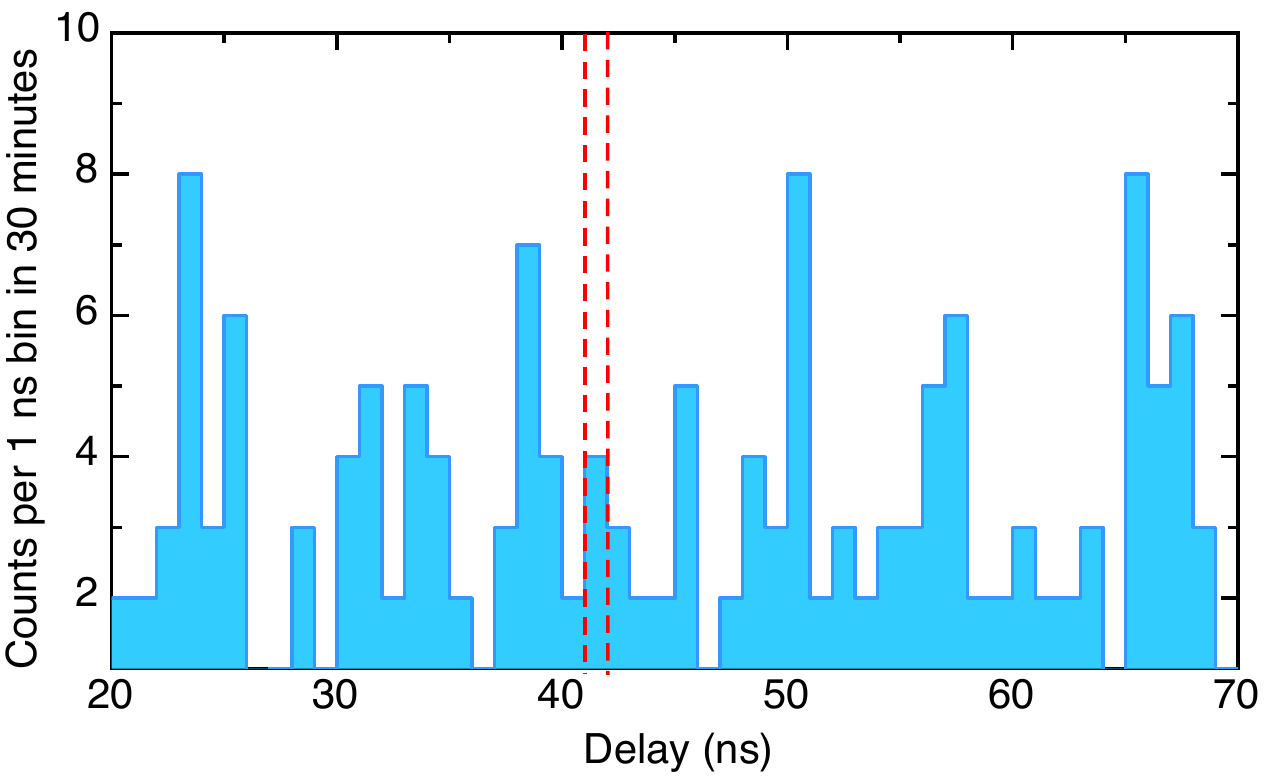}
\caption{Triple coincidences for the detectors H, A, and B, with spacelike separation, measured in a window of $1$ns during a time period of $30 \;$minutes. $R_{HAB}=4$/$30$min. The selected window corresponds to zero delay between clicks at both detectors.}
\label{f4}
\end{figure}

\begin{figure}[htbp]
\includegraphics[width=80 mm]{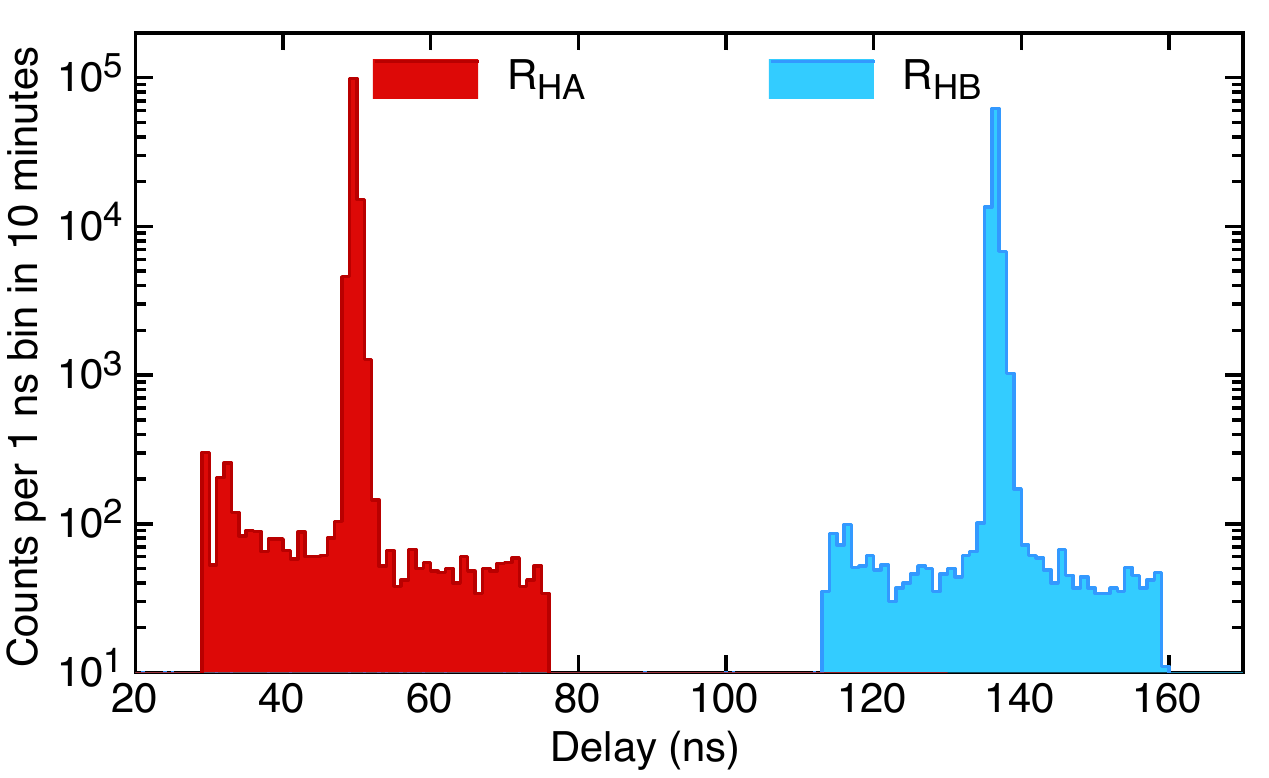}
\caption{Coincidences between the heralding detector and each of the detectors A (red) and B (blue) with timelike separation, measured in a window of \SI{1}{ns} during a time period of 10 minutes. $R_{HA}=$\SI{9.90e4}/10 min, $R_{HB}=$\SI{6.22e4}/10\,min.}
\label{f5}
\end{figure}

\begin{figure}[htbp]
\includegraphics[width=80 mm]{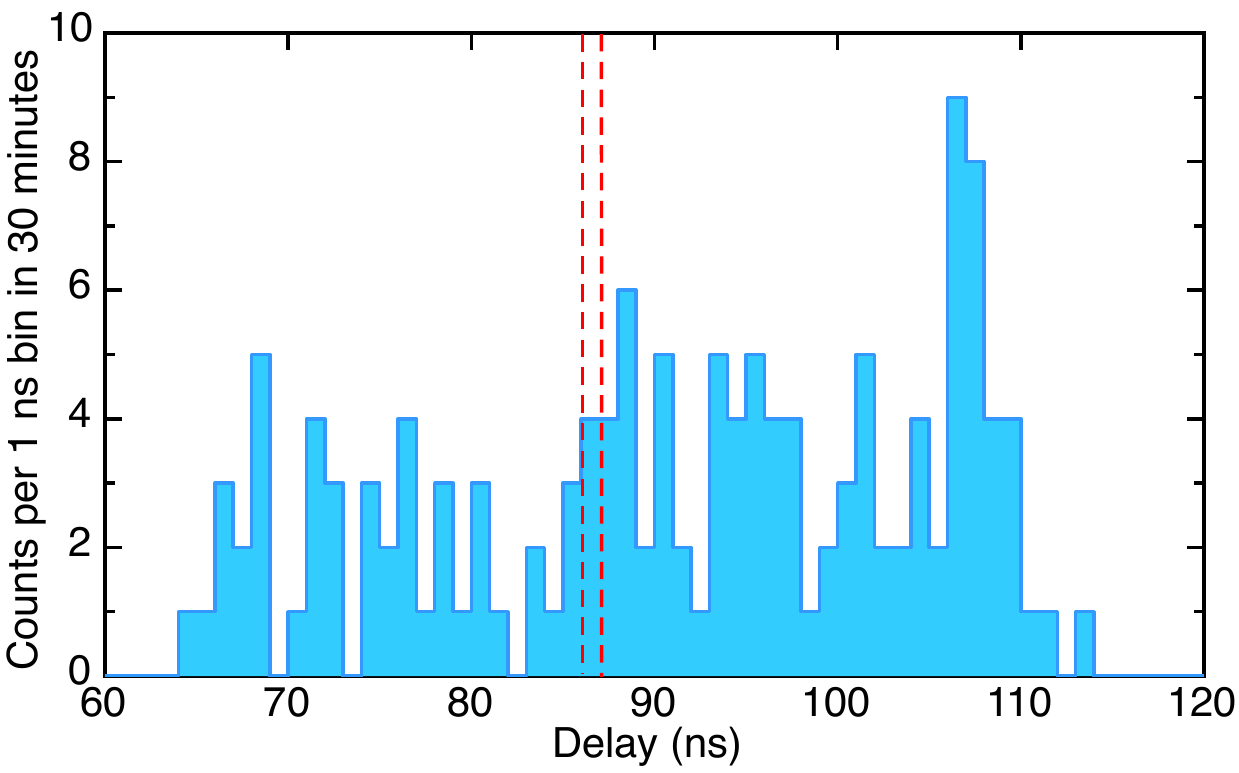}
\caption{Triple coincidences for the detectors H, A, and B, with timelike separation measured in a window of $1$ns during a time period of $30 \;$minutes. $R_{HAB}=4/30$\,min.}
\label{f6}
\end{figure}
First we measure the probabilities of detecting a photon at detector A or at detector B given that a heralding photon has been detected at H. We denote $R_{HA}$ the total number of coincident counts at detector H and detector A during the time of measurement, and $R_{H(A)}$ the total number of counts at detector H alone during the same measurement; $R_{HB}$ and $R_{H(B)}$ denote similar quantities for the measurement with H and B. 

\begin{table}
\begin{center}
\begin{tabular}{@{} c  c  c @{}}
\multicolumn{3}{c}{\textbf{Spacelike separation}}                                                   \\
\hline
\hline
  $R_{HA}$           & $R_{H(A)}$           &  $P^{SL}_A=R_{HA}/R_{H(A)} $                 \\
  $ (94.8 \pm0.3)\cdot10^{3} $   & $(5570  \pm 2 )$ $\cdot 10^{3}$  &  ($1.703 \pm0.006)\cdot 10^{-2}$  \\
  \hline
  $R_{HB}$           & $R_{H(B)}$           &  $P^{SL}_B=R_{HB}/R_{H(B)} $                 \\
  $(63.8  \pm0.2)\cdot10^{3} $   & $(5860  \pm 2) $ $\cdot 10^{3}$  &  $(1.090\pm0.004)\cdot 10^{-2}$  \\
  \hline
  $R_{HAB}$          & $R_{H(AB)}$          &  $P^{SL}(1,1)=R_{HAB}/R_{H(AB)} $            \\
  $4      \pm2$      & $(17145 \pm 4)\cdot 10^{3}$  &  $(2.3 \pm 1.2)\cdot10^{-7}$      \\
  \hline
  $R_{HN}$           & $R_{H(N)}$           &  $P^{SL}_N =R_{HN}/R_{H(N)} $                \\
  $50 \pm 7 $        & $(5500  \pm 2)\cdot10^{3}$  &  $(9.0 \pm1.3)\cdot10^{-6}$         \\
\hline
\multicolumn{3}{c}{}
\\
\multicolumn{3}{c}{\textbf{Timelike separation}}                                                    \\
\hline
\hline
  $R_{HA}$           & $R_{H(A)}$           &  $P^{TL}_A=R_{HA}/R_{H(A)} $                 \\
  $ (99.0 \pm0.3)\cdot10^{3} $   & $(6130 \pm 2)\cdot10^{3}$  &  $(1.616\pm0.005)\cdot 10^{-2} $ \\
  \hline
  $R_{HB}$           & $R_{H(B)}$           &  $P^{TL}_B=R_{HB}/R_{H(B)} $                 \\
  $ (62.2 \pm0.2)\cdot10^{3} $   & $(6100  \pm 2)\cdot10^{3}$  &  $(1.019\pm0.004)\cdot 10^{-2} $ \\
  \hline
  $R_{HAB}$          & $R_{H(AB)}$          &  $P^{TL}(1,1)=R_{HAB}/R_{H(AB)} $            \\
  $4      \pm2$      & $(18345 \pm 4)\cdot10^{3}$  &  $(2.2\pm 1.1)\cdot10^{-7}$     \\
\hline
\end{tabular}
\caption{\label{results}{\bf Summary of results.} Values obtained for the different counting rates and corresponding probabilities defined in the text, measured with spacelike and timelike separation. Statistical errors (one standard deviation) are calculated assuming Poissonian statistics.}
\end{center}
\end{table}

Next we measure the probability of detectors A and B clicking at the same time, again given a heralding signal. $R_{HAB}$ denotes the number of triple coincident counts at the detectors H, A and B, and $R_{H(AB)}$  the total number of counts at detector H alone during the same measurement. All these quantities are measured directly for both a space-like configuration and a time-like configuration. The raw TDC data is shown in Figures~\ref{f3},\ref{f4},\ref{f5} and \ref{f6} and the results are summarized in  Table~\ref{results}.

The number of counts given by detector noise and two-photon events can be estimated by looking at the counts away from the peak.  As an example, for the space-like configuration (Figure \ref{f3}) in a window of 1\,ns the noise rate is on average $R_{HN}=50 $ $(7)$ for a 10 minutes integration time. This corresponds to a noise probability $P_N= 9\cdot~10^{-6} (1.3\cdot10^{-6})$.

From the values in Table 1 one derives the following probability values for spacelike separation:

\begin{eqnarray}
&&P^{SL}_A \cdot P^{SL}_B=1.86 \pm 0.01\cdot10^{-4}\nonumber\\
&&P^{SL}_{AB}=0.002 \pm 0.001\cdot10^{-4};
\label{PABSL}
\end{eqnarray}

\noindent for timelike separation one derives the values:

\begin{eqnarray}
&&P^{TL}_A \cdot P^{TL}_B=1.65 \pm 0.01\cdot10^{-4}\nonumber\\
&&P^{TL}_{AB}=0.002 \pm 0.001\cdot10^{-4};
\label{PABTL}
\end{eqnarray}

\noindent for the probability $P^{SL}_N$ that A and B detect photons coming from different pairs (noise) one derives the value:

\begin{eqnarray}
P^{SL}_N(1,1)&=& P^{SL}_N \cdot P^{SL}_A+P^{SL}_N \cdot P^{SL}_B\nonumber\\
& \approx & 0.0025 \pm 0.0026 \cdot10^{-4}
\label{PN}
\end{eqnarray}

These results show that whether the separation between the detectors is time-like or space-like, the number of coincidences is three orders of magnitude smaller than what would be expected had the events been uncorrelated, i.e., $P_{AB} = P_{A}\times P_{B}$. 

\section{Conclusion}

Einstein was rightly shocked by the claimed completeness of quantum theory \cite{rc}. Indeed, it is surprising that space-like separated detectors can somehow coordinate the detection event of a single photon. Yet, this is obviously necessary to preserve such a fundamental rule as energy conservation. Today, most physicists - if not all~-–~will find our result, space-like anti-bunching of single-photons, as an evidence. Actually, most quantum optics specialists even regard this as the signature of single-photons, though without paying attention to space-like separation. But it might be healthy to keep a feeling of this surprising aspect of quantum physics which reveals a simple form of nonlocality: the conservation of energy in each single event implies nonlocal coordination of detection \cite{as10}. Further considerations on the similarities and differences between the kind of nonlocality demonstrated by our experiment and the better known form of nonlocality revealed by the violation of Bell's inequality will be presented by some of us in a forthcoming publication.

\emph{Acknowledgments}: We are thankful to Ina Domanska for preliminary experimental work. Financial support for this project was provided by the Swiss NCCR Quantum Photonics and by the European Q-ESSENCE project.

\end{document}